\documentclass[letterpaper]{article} % DO NOT CHANGE THIS
\usepackage{aaai2027}  % DO NOT CHANGE THIS

\usepackage[hyphens]{url}  % DO NOT CHANGE THIS
\usepackage{graphicx} % DO NOT CHANGE THIS
\usepackage{natbib}  % DO NOT CHANGE THIS AND DO NOT ADD ANY OPTIONS TO IT
\usepackage{caption} % DO NOT CHANGE THIS AND DO NOT ADD ANY OPTIONS TO IT
\usepackage{algorithm}
\usepackage{algorithmic}
\usepackage{url}
\usepackage{latexsym}
\usepackage{amsmath}
\usepackage{amssymb}
\usepackage{booktabs}
\usepackage{multirow}
\usepackage[table]{xcolor}
\usepackage{tcolorbox}
\usepackage{xcolor}
\usepackage{tikz}
\usepackage{mdframed}
\usepackage{makecell}
\usepackage{graphicx}
\newcommand{\modelname}{TALRanker}

\usepackage{newfloat}
\usepackage{listings}
\DeclareCaptionStyle{ruled}{labelfont=normalfont,labelsep=colon,strut=off} % DO NOT CHANGE THIS
\floatstyle{ruled}
\newfloat{listing}{tb}{lst}{}
\floatname{listing}{Listing}

\usepackage{booktabs}

\title{Tool-Adaptive LLM Reranker}
\author {
    Zichuan Liu\textsuperscript{\rm 1}\equalcontrib\corresponding,
    Ruijin Hua\textsuperscript{\rm 2}\equalcontrib
}
\affiliations {
    \textsuperscript{\rm 1}Carnegie Mellon University, Pittsburgh, PA 15213, USA\\
    \textsuperscript{\rm 2}Huazhong University of Science and Technology, Wuhan, China\\
    zichuanl@andrew.cmu.edu, ruijinhua07@gmail.com
}

\begin{document}

\maketitle

\begin{abstract}
Generative Large Language Models (LLMs) have revolutionized information retrieval, yet their strictly parametric nature frequently leads to severe factual hallucinations when confronted with complex queries beyond their epistemic boundaries. 
While external tool-calling can mitigate this, indiscriminately invoking search tools for every document during reranking incurs prohibitive latency overheads, creating an intractable accuracy-efficiency dilemma. 
To address this challenge, we propose \modelname, a novel framework that formalizes pointwise relevance scoring as an agentic Markov decision process. 
We optimize it via a two-stage training paradigm. 
An initial warm-up utilizes a language-preserving hybrid loss to prevent the catastrophic forgetting of native generative capacities. 
Subsequently, an asymmetric cost-aware reward equipped in reinforcement learning forces the policy to autonomously bypass tools for maximum efficiency when confident, while selectively retrieving external evidence to avert severe hallucination penalties when uncertain. 
Extensive evaluations demonstrate that \modelname~achieves state-of-the-art performance across standard and reasoning-intensive retrieval benchmarks, matching throughput with pointwise rerankers while outperforming parameter-heavy reasoning models.
\end{abstract}

\section{Introduction}

Driven by the escalating complexity of user needs, information retrieval systems are undergoing a fundamental shift: from executing simple keyword searches to resolving complex, inferential questions that demand deep reasoning~\cite{weller-etal-2025-followir,mao-etal-2024-chatretriever,Guo_2025}. In advanced applications such as Retrieval-Augmented Generation (RAG) and autonomous agents~\cite{yu2024rankrag,li2026semanticsimilarityrethinkingretrieval,10.1145/3770854.3783917,wang2024rcagent}, accurately assessing relevance under these inferential scenarios is critical. Consequently, traditional scalar models are largely superseded by Generative Relevance Models (GRMs) powered by Large Language Models (LLMs)~\cite{li2025matchinggenerationsurveygenerative}. Specifically, reasoning-based GRMs that leverage chain-of-thought~\cite{wei2022chain} have demonstrated remarkable capabilities in analyzing query-document relationships~\cite{zhang2025rearankreasoningrerankingagent,weller2025rank}. 
However, these models are fundamentally constrained by their parametric boundaries, operating in a ``closed-book" setting, and the internal reasoning pathways inevitably break down when confronted with out-of-distribution (OOD) queries, long-tail knowledge, or highly specific nuances. Exceeding these epistemic boundaries frequently leads to severe factual hallucinations and overconfident, erroneous relevance scores.

\begin{figure}[t]
    \centering
    \includegraphics[width=\columnwidth]{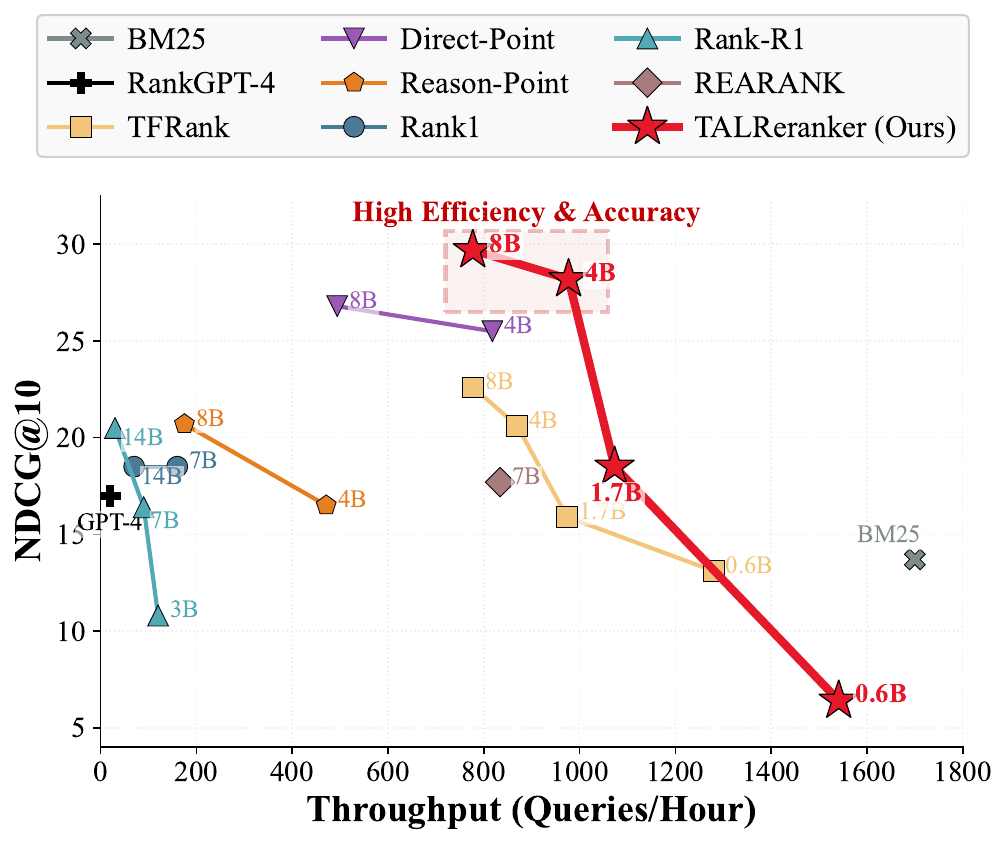}
    \caption{The tradeoff between precision (NDCG@10) and single-GPU throughput on the BRIGHT benchmark. \modelname~dominates the top-right quadrant, achieving state-of-the-art accuracy while maintaining extreme inference speeds comparable to non-reasoning discriminative baselines.}
    \label{fig:efficiency}
\end{figure}

In the broader landscape of LLM agents, external knowledge retrieval via tool-calling emerges as a proven mechanism to mitigate hallucinations~\cite{yang2026efficientagentsmemorytool,qian-etal-2025-smart,singh2025agentic,schick2023toolformer,yao2023reactsynergizingreasoningacting}. Nevertheless, effectively integrating this paradigm into the reranking system remains largely unresolved. Existing approaches either operate as purely parametric models that obstinately abstain from tool use~\cite{zhuang2025rankr1enhancingreasoningllmbased,zhang2025rearankreasoningrerankingagent,lu2026rethinking}, or naively trigger agentic search with tools for every single query~\cite{gou2024critic,chern2025factool}. 
Given that reranking in real enterprise-level search engines requires evaluating hundreds of query-document pairs per session, unconditional tool-calling incurs prohibitive computational overhead, severely violating the latency budgets of practical applications~\cite{yoon2025acurank}.
Furthermore, recent attempts to accelerate inference by completely internalizing reasoning~\cite{fan2025tfrankthinkfreereasoningenables} fundamentally fail to resolve the lack of external grounding when critical factual augmentation is required.

In this work, we consider that the core challenge lies not merely in enabling tool use, but in orchestrating a cost-sensitive, adaptive routing mechanism. 
To this end, we formulate pointwise relevance estimation as an agentic Markov decision process and propose the \underline{\textbf{T}}ool-\underline{\textbf{A}}daptive \underline{\textbf{L}}LM \underline{\textbf{R}}er\underline{\textbf{anker}} (\modelname). 
Instead of relying on static scoring or forced external retrieval, \modelname~initiates an explicit reasoning trajectory to dynamically assess its internal knowledge confidence. When parametric certainty is high, the model directly emits a relevance score, bypassing tools to minimize latency. Conversely, upon detecting an internal knowledge deficit, it actively suspends generation to trigger an external tool, grounding its final judgment in the retrieved evidence. 
Crucially, we optimize \modelname~through two-stage training. First, an initial warm-up stage applies a language-preserving hybrid loss that equips the model with discriminative scoring abilities while preventing the catastrophic forgetting of its native generative capacities. Subsequently, in the reinforcement learning stage, we adopt GRPO~\cite{Guo_2025} equipped with an asymmetric cost-aware reward mechanism. 
In most cases, this forces bypassing tools to achieve maximum efficiency when confident, but pays controlled delay penalties to retrieve external knowledge and avoid severe hallucination penalties when uncertain.
As illustrated in Figure~\ref{fig:efficiency}, this tool-adaptive strategy empowers \modelname{} to dominate the accuracy-efficiency trade-off, making it achieve state-of-the-art NDCG on the challenging benchmark while providing fast inference.

\begin{figure*}[t]
    \centering
    \includegraphics[width=0.8\textwidth]{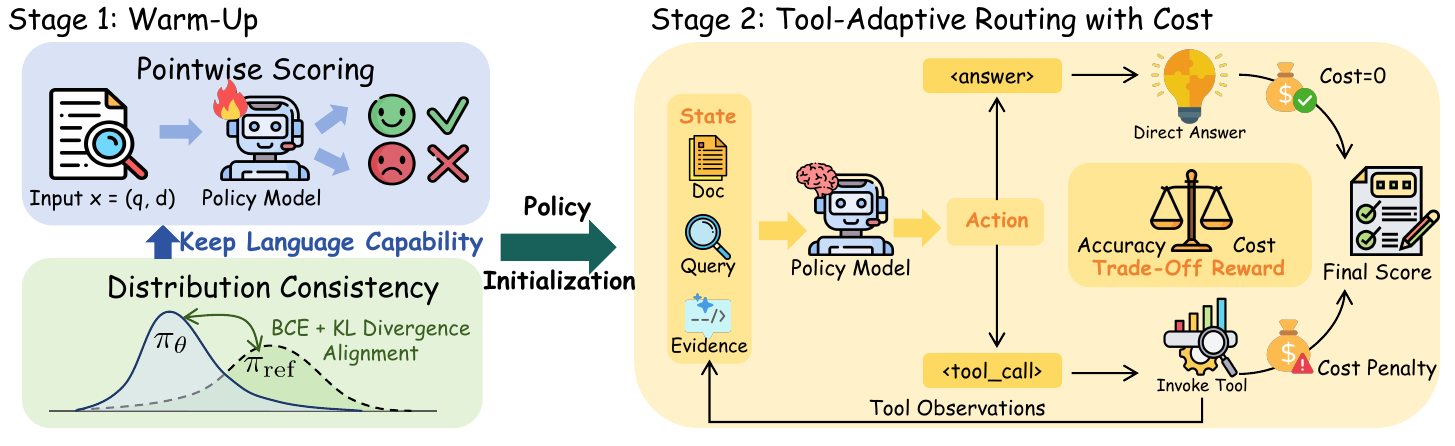}
    \caption{The overall training framework of \modelname. Stage 1 performs a warm-up utilizing a language-preserving hybrid loss to establish scoring capabilities while preventing catastrophic forgetting. Stage 2 formulates the generative reranking as an agentic MDP, employing a cost-aware reward mechanism to dynamically balance accuracy against tool invocation latency.}
    \label{fig:main}
\end{figure*}

Our key contributions are summarized as follows:
\begin{itemize}
    \item We propose \modelname, a novel pointwise reranking paradigm that formalizes relevance estimation as an agentic Markov decision process. It seamlessly bridges discriminative scoring with generative tool-calling to autonomously manage its epistemic boundaries.
    \item We introduce a robust two-stage training framework featuring a language-preserving hybrid loss and an asymmetric cost-aware reward. This design prevents catastrophic capability degradation and explicitly forces the policy to master the optimal tradeoff between latency and accuracy.
    \item Extensive evaluations demonstrate that \modelname~dominates the accuracy-efficiency tradeoff. It establishes a new state-of-the-art across standard and reasoning-intensive benchmarks, matching the extreme throughput of non-reasoning baselines while significantly outperforming parameter-heavy reasoning models.
\end{itemize}
\section{Related Works}
\textbf{Generative Relevance Models.} The application of large language models in relevance ranking primarily encompasses three paradigms: pointwise~\cite{li2026prorankpromptwarmupreinforcement,sun2023is,liang2023holistic}, pairwise~\cite{qin-etal-2024-large,luo-etal-2024-prp}, and listwise methods~\cite{xu2026beyond,Zhuang_2024}. In large-scale industrial search systems, evaluating a massive volume of candidate documents is mandatory~\cite{10.1145/3770854.3783917,zhang2025qwen3embeddingadvancingtext,li2026qwen3vlembeddingqwen3vlrerankerunifiedframework}. Pairwise and listwise methods incur prohibitive computational costs under such conditions. Consequently, pointwise reranking establishes itself as the dominant paradigm due to its highly parallelizable architecture and $\mathcal{O}(1)$ inference complexity per query-document pair~\cite{li2025matchinggenerationsurveygenerative}. 
Recent advancements attempt to augment pointwise rerankers by integrating explicit reasoning trajectories~\cite{zhuang2025rankr1enhancingreasoningllmbased}, compression mechanisms~\cite{fan2025tfrankthinkfreereasoningenables}, and initial agentic behaviors~\cite{liu2026reasonrankempoweringpassageranking,zhang2025rearankreasoningrerankingagent,li2026semanticsimilarityrethinkingretrieval}. However, these approaches share a critical bottleneck: they operate strictly within a parametric boundary. Lacking the mechanism to access external knowledge, these ``closed-book" reasoning models inevitably suffer from factual hallucinations and degraded accuracy when confronted with out-of-distribution or long-tail queries.

\textbf{Efficient Tool-Augmented Agents.} Equipping LLMs with external tool-calling capabilities significantly expands their operational boundaries and mitigates parametric knowledge limitations~\cite{yao2023reactsynergizingreasoningacting,wang2024rcagent}. Because tool-calling inherently introduces severe latency, recent literature focuses heavily on optimizing execution efficiency~\cite{yang2026efficientagentsmemorytool}. Notable efforts include refining tool selection mechanisms~\cite{qu2025from,erdogan-etal-2024-tinyagent}, eliminating redundant invocations~\cite{gao-etal-2025-efficient,su2025toolorchestraelevatingintelligenceefficient}, and synthesizing tool usage with reasoning pathways via reinforcement learning~\cite{qian-etal-2025-smart,liu2025sample}. 
Despite these advances in general-purpose agents, deploying efficient tool usage within pointwise reranking presents a profound contradiction. Strict latency budgets preclude reasoning and tool-using execution across massive document pools~\cite{fan2025tfrankthinkfreereasoningenables}. Conversely, enforcing purely parametric evaluation on complex queries induces factual hallucinations~\cite{zhuang2025rankr1enhancingreasoningllmbased,zhang2025rearankreasoningrerankingagent}. Existing frameworks either abstain from tool usage entirely or struggle to adapt generation without explicit search interventions~\cite{weller-etal-2025-followir}. This highlights a critical void for an adaptive routing strategy for the computational demands of relevance reranking. To bridge this gap, our proposed method elegantly combines discriminative scoring capabilities with highly efficient, generative tool-calling mechanisms to effectively resolve this contradiction.

\section{Problem Formulation}
\label{sec:problem_formulation}

Given a user query $q$ and a set of candidate documents $\mathcal{D} = \{d_1, d_2, \dots, d_n\}$ retrieved by a coarse-grained passage retriever, the goal of standard pointwise reranking is to estimate a relevance score $s_i$ for each isolated $(q, d_i)$ pair to determine the final ranked permutation. In LLM-based reranking, this is typically formulated as a ``closed-book" generative or scoring process. However, forcing the model to rely exclusively on its internal parametric memory frequently leads to factual hallucinations, especially when confronted with out-of-distribution queries or complex demands.

To mitigate hallucinated scoring while preserving the highly parallelizable $\mathcal{O}(1)$ complexity of pointwise ranking, we reformulate relevance estimation as an \textit{agentic Markov decision process}. In this paradigm, rather than performing static, single-pass scoring, the generative reranker $\pi_\theta$ dynamically controls the computational pathway. 
Specifically, for each document $d_i$, $\pi_\theta$ first generates an explicit reasoning trajectory $\tau$. If an internal knowledge deficit is detected, the model autonomously triggers an external search tool to retrieve supportive evidence $e$. Upon acquiring sufficient information, the generation terminates with a discrete binary judgment (e.g., \texttt{yes} or \texttt{no}). The final continuous relevance score $s_i$ is then directly derived from the model's terminal predictive probability distribution over these target tokens. Consequently, the core problem of our work is formulated as optimizing this policy $\pi_\theta$ to adaptively balance maximum ranking accuracy against the minimal computational latency incurred by tool usage.

\section{Methodology}
\label{sec:methodology}

In this section, we detail the proposed \underline{\textbf{T}}ool-\underline{\textbf{A}}daptive \underline{\textbf{L}}LM \underline{\textbf{R}}er\underline{\textbf{anker}} (\modelname), the overall framework is illustrated in Figure~\ref{fig:main}. We first establish the foundational mechanism of generative pointwise scoring, highlighting the necessity of preserving the native language capabilities of the model. Subsequently, we formalize the search-augmented reranking process as a discrete-state Markov Decision Process (MDP). Finally, we introduce the cost-aware reinforcement learning framework, elucidating how the asymmetric penalty design drives the generative policy to balance reasoning accuracy against execution latency.

\subsection{Language-Preserved Discriminative Scoring}

Given a user query $q$ and a candidate document $d$, a Large Language Model (LLM) performs relevance estimation by processing a constructed prompt $x = \text{Prompt}(q, d)$ that incorporates system instructions. The final continuous relevance score $s$ is extracted by normalizing the predictive probabilities over specific target tokens (i.e., \texttt{yes} and \texttt{no}):
\begin{equation}
    \label{eq:score}
    s = \frac{\pi_\theta(\text{\texttt{yes}} | x)}{\pi_\theta(\text{\texttt{yes}} | x) + \pi_\theta(\text{\texttt{no}} | x)}.
\end{equation}

In standard supervised fine-tuning, adapting the model to this binary discriminative paradigm typically relies exclusively on Binary Cross-Entropy (BCE) optimization. However, exclusively minimizing the discriminative loss triggers catastrophic forgetting. This degradation reduces the LLM into a rigid scalar function and irreversibly strips its ability to generate complex reasoning trajectories or invoke external tools.
% To prevent this capability degradation, we introduce a hybrid regularized objective during the warmup stage. Specifically, we augment the traditional binary loss $\mathcal{L}_{\text{CE}}$ with a KL divergence penalty relative to the original base model $\pi_{\text{ref}}$:
% \begin{equation}
%     \mathcal{L}_{\text{warmup}} = \mathcal{L}_{\text{CE}}(s, s_{\text{gt}}) + \beta \mathbb{D}_{\text{KL}}(\pi_\theta || \pi_{\text{ref}}),
% \end{equation}
% where $s_{\text{gt}}$ represents the ground-truth binary relevance label. The hyperparameter $\beta$ controls the intensity of preserving the generative capacity of the model. 
While applying KL divergence penalties is a standard practice to prevent policy collapse, pointwise discriminative scoring requires a specialized adaptation. If applied uniformly, the KL constraint would prevent the model from confidently predicting the highly deterministic \texttt{<yes>} or \texttt{<no>} target tokens. Therefore, we design a \textit{masked} hybrid objective. 
Specifically, for a sequence of length $T$, we apply the binary cross-entropy loss $\mathcal{L}_{\text{BCE}}$ exclusively at the terminal scoring token. Simultaneously, we apply a token-wise KL penalty over the entire vocabulary $\mathcal{V}$ for all preceding context tokens $x_{<T}$, anchoring them to the reference model $\pi_{\text{ref}}$:
\begin{equation}
\begin{split}
    \mathcal{L}_{\text{warmup}} &= \mathcal{L}_{\text{BCE}}(s, s_{\text{gt}}) \\
    &\quad + \frac{\beta}{T-1} \sum_{t=1}^{T-1} \mathbb{D}_{\text{KL}} \big( \pi_{\text{ref}}(\cdot | x_{<t}) \,\|\, \pi_\theta(\cdot | x_{<t}) \big),
\end{split}
\end{equation}
where $s_{\text{gt}}$ represents the ground-truth relevance label and $\beta$ controls the regularization intensity. By explicitly masking out the final scoring token from the KL penalty, the policy equips the policy network with proficient discriminative scoring capabilities, while remaining strictly anchored to the foundational language distribution across all preceding reasoning trajectories for executing sophisticated, tool-augmented explorations in subsequent stages.

\subsection{Tool-Adaptive Reranking via MDP}
To achieve fine-grained control and optimization over tool invocations, we reformulate the generative reranking process as a token-level MDP, denoted by the tuple $(\mathcal{S}, \mathcal{A}, \mathcal{P}, \mathcal{R})$.

\textbf{State and Action Space.} A state $h_t \in \mathcal{S}$ is defined as the complete sequential context at time step $t$, encompassing the initial prompt $x$, the generated reasoning trajectory, and any external tool observations. The action space $\mathcal{A}$ constitutes the vocabulary $\mathcal{V}$ of the LLM. Crucially, $\mathcal{A}$ includes two distinct special actions (target tokens) that dictate the computational graph: \texttt{<tool\_call>} and \texttt{<answer>}.

\textbf{Transition Dynamics.} The reranker, acting as the policy $\pi_\theta(a_t | h_t)$, samples actions autoregressively. For standard vocabulary tokens, the environment simply appends them to the current state to continue the internal reasoning process. Crucially, when the policy emits a \texttt{<tool\_call>} action, the generative process is temporarily suspended. The environment intercepts this action, executes an external retrieval operation to fetch supportive evidence $e$, and deterministically updates the state by appending this observation (i.e., $h_{t+1} = h_t \oplus e$). This transition mechanism effectively bridges the parametric model with external knowledge sources. Following the observation, the model resumes its generative exploration. Specifically, the policy typically generates a sequence of standard vocabulary tokens enclosed within \texttt{<reasoning>} tags to internally digest and deduce over the assimilated evidence. This chain-of-thought sequence iterates until the policy emits the terminal \texttt{<answer>} action. This specific action triggers the trajectory to cease, prompting the model to evaluate the accumulated context and project its final hidden representation into the continuous relevance score $s$ via Eq.~(\ref{eq:score}). We define the entire generated sequence of states and actions from the initial prompt to this terminal action as a complete trajectory $\tau$.

This formalization breaks the traditional ``closed-book" rerankers. It establishes an alternating cycle of generative exploration and terminal discrimination, forging a rigorous framework for subsequent reinforcement-based reward allocation $\mathcal{R}$. Notably, this state transition mechanism is uniformly executed during the inference stage, where the model autoregressively interacts with the search environment until termination. We provide the detailed algorithmic execution flow for inference in the Appendix.

\subsection{Cost-Aware Reward Design in RL}

Having established the MDP framework, our ultimate objective is to optimize the policy $\pi_\theta$ to adaptively trigger external tools upon reaching its parametric knowledge boundaries, thereby balancing reranking accuracy and execution latency. To achieve this, we adopt a cost-aware reward with the Group Relative Policy Optimization (GRPO)~\cite{Guo_2025}.

\textbf{Trajectory Sampling.} During each training iteration, given a prompt $x$, the current policy $\pi_\theta$ first autoregressively samples a group of $G$ complete decision trajectories, denoted as $\{\tau_1, \tau_2, \dots, \tau_G\}$. Subsequently, we evaluate the return for each trajectory and update the policy by relying on the relative advantages within the group. 
For a generated trajectory $\tau$ within the sampled group, we formulate a comprehensive scalar reward $\mathcal{R}(\tau)$. This reward explicitly evaluates format compliance, predictive accuracy, and the computational cost of tool-calling through three components:
\begin{equation}
    \mathcal{R}(\tau) = \mathcal{R}_{\text{format}} + \mathcal{R}_{\text{acc}} + \mathcal{R}_{\text{cost}}.
\end{equation}

\textbf{Format Verification.} The term $\mathcal{R}_{\text{format}}$ enforces strict structural compliance. It yields a binary reward of $1.0$ only if the generated trajectory strictly adheres to predefined valid patterns (i.e., a zero-hop direct terminal action, or a multi-hop reasoning sequence concluding validly with \texttt{<answer>}); otherwise, it yields $0.0$.

\textbf{Execution Cost Penalty.} To penalize unconditional searching and ensure industrial viability, we impose a  latency penalty for every tool invoked during the trajectory:
\begin{equation}
    \mathcal{R}_{\text{cost}} = -\lambda \cdot \mathcal{N}_{\text{hops}},
\end{equation}
where $\mathcal{N}_{\text{hops}}$ denotes the total number of tool calls, and $\lambda$ acts as the scaling coefficient for the penalty.

\textbf{Asymmetric Accuracy Reward.} This component is the pivotal driver of adaptive behavior. Given a tolerance threshold $\gamma$, we evaluate the absolute prediction error $|s - s_{\text{gt}}|$ between the final relevance score $s$ (derived by $\pi_\theta$ at the terminal state) and the ground-truth label $s_{\text{gt}}$. We formulate an asymmetric piecewise reward:
\begin{equation}
\mathcal{R}_{\text{acc}} = 
\begin{cases} 
    1.0 - |s - s_{\text{gt}}|, & \text{if } |s - s_{\text{gt}}| \le \gamma \\
    -1.0, & \text{if } |s - s_{\text{gt}}| > \gamma \text{ and } \mathcal{N}_{\text{hops}} = 0 \\
    0.0, & \text{if } |s - s_{\text{gt}}| > \gamma \text{ and } \mathcal{N}_{\text{hops}} > 0
\end{cases}.
\end{equation}
This asymmetric reward structurally coerces the model to internalize its epistemic uncertainty. 
% When internal knowledge is deficient, forcing a zero-hop prediction risks a severe catastrophic penalty ($-1.0$). Conversely, while invoking a tool incurs a deterministic computational cost (via $\mathcal{R}_{\text{cost}}$), it effectively safeguards the policy from the severe penalty of hallucinated failures by neutralizing the accuracy penalty to $0.0$. 
When encountering internal knowledge deficits, invoking a tool incurs a minor deterministic cost but effectively safeguards the policy from the severe catastrophic penalty ($-1.0$) associated with hallucinatory zero-hop predictions.
Conversely, when the model answers correctly, bypassing tools yields a maximum reward of approximately $1.0$, whereas invoking tools reduces the net reward by the latency penalty $-\lambda \cdot \mathcal{N}_{\text{hops}}$.
Consequently, the policy network $\pi_\theta$ is mathematically driven to dynamically route queries to external tools only when strictly necessary, thereby elegantly resolving the contradiction between ranking accuracy and latency budgets.

\section{Experimental Setup}
\label{sec:settings}
\begin{table*}[t]
\centering
\caption{Evaluation results (NDCG@10) across In-Domain and OOD benchmarks. $^\dagger$ are from the official repositories.}
\label{tab:comprehensive_main_v5}
\resizebox{\textwidth}{!}{
\begin{tabular}{l l | c c | c c c c c c c c c c }
\toprule
\multirow{2}{*}{\textbf{Model}} & \multirow{2}{*}{\textbf{Training}} & \multicolumn{2}{c|}{\textbf{In-Domain}} & \multicolumn{10}{c}{\textbf{Out-of-Domain}} \\
\cmidrule(lr){3-4} \cmidrule(lr){5-14}
& & DL19 & DL20 & ArguA & ClimF & DBPedia & FiQA & NFCorp & SciDoc & SciFact & Touche & TrecC & BEIR (Avg) \\
\midrule
BM25$^\dagger$ & - & 50.6 & 48.0 & 39.7 & 16.5 & 31.8 & 23.6 & 33.8 & 14.9 & 67.9 & 44.2 & 59.5 & 36.9 \\
RankGPT-4$^\dagger$ & zero-shot & 75.6 & 70.6 & - & - & 47.1 & - & 38.5 & - & 75.0 & 38.6 & 85.5 & - \\
setwise.bubblesort (Flan-t5-3B)$^\dagger$ & zero-shot & 70.5 & 67.6 & - & - & 39.7 & 41.2 & - & - & - & - & 71.1 & -  \\
setwise.bubblesort (Flan-t5-11B)$^\dagger$ & zero-shot & 71.1 & 68.6 & - & - & 41.7 & 43.3 & - & - & - & - & 74.9 & -  \\
\midrule
Rank1-7B$^\dagger$ & SFT & - & - & 26.4 & 16.2 & 37.7 & 38.4 & 37.9 & 16.5 & 76.1 & 24.5 & 79.5 & 39.2  \\
Rank1-14B$^\dagger$ & SFT & - & - & 32.2 & 15.6 & 34.2 & 36.6 & 35.1 & 16.6 & 73.8 & 25.9 & 78.0 & 38.7  \\
Reason-Point-4B & SFT & 21.2 & 17.4 & 0.0 & 1.0 & 6.9 & 2.3 & 6.8 & 1.7 & 3.3 & 10.1 & 32.1 & 7.1 \\
Direct-Point-4B & SFT & 66.1 & 54.0 & 59.1 & 20.0 & 44.1 & 42.9 & 37.2 & 17.6 & 75.5 & 23.3 & 81.3 & 44.5 \\
Reason-Point-8B & SFT & 58.6 & 49.0 & 56.3 & 17.8 & 41.7 & 41.3 & 37.3 & 15.8 & 72.6 & 27.3 & 80.0 & 43.3 \\
Direct-Point-8B & SFT &  64.9 & 53.2 & 47.5 & 17.9 & 42.3 & 41.9 & 37.4 & 18.2 & 73.1 & 21.0 & 80.6 & 42.2 \\
Reason-List-4B$^\dagger$ & SFT & 70.8 & 68.7 & 39.3 & 13.8 & 37.7 & 32.9 & 30.2 & 16.0 & 69.1 & 24.6 & 79.5 & 38.1 \\
Direct-List-4B$^\dagger$ & SFT & 73.8 & 69.0 & 41.7 & 14.1 & 36.5 & 37.0 & 33.5 & 16.4 & 72.4 & 22.9 & 78.1 & 39.2 \\
Reason-List-8B$^\dagger$ & SFT & 72.6 & 69.8 & 79.2 & 27.1 & 69.7 & 18.1 & 36.0 & 36.6 & 42.2 & 16.2 & 32.8 & 39.8 \\
Direct-List-8B$^\dagger$ & SFT &73.0 & 71.4 & 80.1 & 27.0 & 73.6 & 18.9 & 35.9 & 37.7 & 42.3 & 16.4 & 28.9 & 40.1 \\
\midrule
TFRank-4B$^\dagger$ & SFT+GRPO & - & - & 37.2 & 19.7 & 37.9 & 36.2 & 38.3 & 18.3 & 76.6 & 37.6 & 80.5 & 42.5 \\
TFRank-8B$^\dagger$ & SFT+GRPO & - & - & 36.5 & 21.7 & 37.0 & 38.0 & 38.0 & 17.9 & 74.6 & 35.0 & 80.0 & 42.1 \\
ReasonRank-7B$^\dagger$ & SFT+GRPO & - & - & 79.6 & 30.4 & 72.8 & 19.7 & 36.6 & 38.2 & 44.7 & 20.0 & 33.3 & 41.7 \\
\midrule
REARANK-7B$^\dagger$ & GRPO & 74.2 & 70.0 & 35.6 & 20.6 & 43.5 & 35.8 & 37.9 & 19.2 & 71.9 & 40.2 & 80.1 & 42.8 \\
Rank-R1-7B$^\dagger$ & GRPO & 72.7 & 68.5 & 37.0 & 24.1 & 43.2 & 40.1 & 36.2 & 18.8 & 76.1 & 33.0 & 82.6 & 43.5 \\
Rank-R1-14B$^\dagger$ & GRPO & 71.4 & 69.1 & 34.4 & 24.2 & 44.0 & 43.0 & 37.9 & 19.7 & 77.5 & 29.6 & 83.9 & 43.8 \\
\midrule
\rowcolor{gray!15} \textbf{TALReranker}-4B & {SFT+GRPO} & \underline{76.6} & \underline{74.7} & {55.5} & {25.6} & {44.9} & {43.6} & {37.9} & {18.8} & {77.8} & {29.6} & {80.4} & \underline{46.0} \\
\rowcolor{gray!15} \textbf{TALReranker}-8B & {SFT+GRPO} & \textbf{77.4} & \textbf{76.6} & {58.4} & {25.6} & {45.6} & {44.5} & {37.4} & {18.9} & {77.3} & {29.6} & {79.4} & \textbf{46.3} \\
\bottomrule
\end{tabular}
}
\end{table*}
\paragraph{Implementation Details}
\modelname{} is built upon the Qwen3 series~\cite{yang2025qwen3}. The training regimen adheres strictly to our proposed two-stage framework. For the initial warmup stage, we fine-tune the model on binary relevance pairs derived from the MS MARCO passage dataset\footnote{Tevatron/msmarco-passage} to establish a robust discriminative foundation. Subsequently, for the GRPO stage, we randomly select a subset of 10,000 samples from the dataset. On this subset, the policy network explores agentic trajectories and is optimized by our latency-aware reward to master the adaptive tool-calling mechanism. The tool uses the Perplexity API. All experiments are executed on an 8-GPU NVIDIA H100 cluster, leveraging the vLLM engine to ensure high-throughput inference. A detailed implementation is in the Appendix.

\paragraph{Evaluation Benchmarks}
To rigorously assess the retrieval efficacy and zero-shot transferability of \modelname{}, we construct a comprehensive evaluation spanning both seen and unseen distributions. For in-domain assessment, we employ TREC-DL19~\cite{craswell2020overviewtrec2019deep} and TREC-DL20~\cite{craswell2021overviewtrec2020deep}. For Out-of-Domain~(OOD) scenarios, we curate a diverse subset of 9 datasets from the BEIR benchmark~\cite{thakur2021beir}, encompassing specialized domains such as biomedical text (TREC-COVID, NFCorpus), fact verification (SciFact, Climate-FEVER), and financial documents (FiQA). Furthermore, to explicitly stress-test the model's capacity for complex logic, we utilize the BRIGHT benchmark~\cite{su2025bright}, consisting of 1,384 challenging, multi-hop reasoning queries across coding, mathematics, and economics. Following prior
work, model performance is consistently quantified using Normalized Discounted Cumulative Gain at 10 (NDCG@10).

\paragraph{Baselines}
We evaluate \modelname{} against a diverse set of ranking methods covering different modeling paradigms and training regimes. As a traditional lexical method, we include BM25~\cite{10.1145/3404835.3463238} in a direct inference setting. We further consider LLM-based rerankers without task-specific training, represented by RankGPT~\cite{sun2023is}.
For supervised approaches, we include Rank1~\cite{weller2025rank} and TFRank~\cite{fan2025tfrankthinkfreereasoningenables}, which learn ranking functions via fine-tuning on labeled or synthetic data, alongside Flan-t5~\cite{Zhuang_2024} configured for setwise ranking. In addition, we evaluate reinforcement learning-based methods, including Rank-R1~\cite{zhuang2025rankr1enhancingreasoningllmbased} trained with GRPO, REARANK-7B~\cite{zhang2025rearankreasoningrerankingagent}, which models reranking as a multi-step decision process; and ReasonRank~\cite{liu2026reasonrankempoweringpassageranking}, which elicits reasoning paths for relevance. 

\section{Results and Analysis}

\begin{table*}[t!]
\centering
\caption{Adaptive tool invocation ratio (\%) of \modelname~across the TREC and BRIGHT benchmark subsets.}
\label{tab:adaptive_hops}
\resizebox{\textwidth}{!}{
\begin{tabular}{l | c c | c c c c c c c c c c c c}
\toprule
\multirow{2}{*}{\textbf{Scale}} & \multicolumn{2}{c|}{\textbf{DL}} & \multicolumn{12}{c}{\textbf{BRIGHT}} \\
\cmidrule(lr){2-3} \cmidrule(lr){4-15}
& DL19 & DL20 & Bio. & Earth. & Econ. & Psy. & Rob. & Stack. & Sus. & Pony & Leet. & AoPS & TheoT. & TheoQ. \\
\midrule
0.6B & 100.00 & 100.00 & 100.00 & 100.00 & 100.00 & 100.00 & 100.00 & 100.00 & 100.00 & 100.00 & 100.00 & 100.00 & 100.00 & 100.00 \\
1.7B & 0.04 & 0.10 & 0.37 & 0.47 & 0.41 & 2.08 & 0.67 & 0.30 & 0.45 & 1.47 & 0.06 & 0.11 & 0.42 & 0.09 \\
4B & 0.00 & 0.03 & 0.15 & 0.19 & 0.27 & 1.25 & 0.31 & 0.14 & 0.23 & 0.82 & 0.00 & 0.00 & 0.12 & 0.03 \\
8B & 0.00 & 0.00 & 0.02 & 0.05 & 0.05 & 0.28 & 0.04 & 0.02 & 0.04 & 0.11 & 0.00 & 0.00 & 0.02 & 0.00 \\
\bottomrule
\end{tabular}
}
\end{table*}

\subsection{Main Results}
We evaluate the proposed \modelname~against a comprehensive suite of baseline models across standard and reasoning-intensive retrieval benchmarks. The empirical results demonstrate the superior effectiveness and efficiency of our method.

\textbf{Performance on Standard Retrieval Benchmarks.} 
We report the NDCG@10 performance on the in-domain TREC tracks and OOD datasets from the BEIR benchmark in Table~\ref{tab:comprehensive_main_v5}. Overall, \modelname-8B establishes the best performance in both settings, achieving a remarkable new state-of-the-art. Notably, our 4B variant yields an average BEIR score of 46.0, directly surpassing heavily scaled models such as Rank-R1-14B and Rank1-14B, and this superiority consistently extends to the in-domain evaluations. This indicates that empowering the model with adaptive tool invocation is vastly more efficient than purely scaling up the underlying parametric memory. Furthermore, unlike traditional pointwise and listwise approaches, our cost-aware RL enables \modelname~to avoid rigid sequence generation while maintaining high accuracy across diverse datasets.

\textbf{Performance on Reasoning-Intensive Benchmarks.} 
As illustrated in Table~\ref{tab:bright_detailed_evaluation}, we evaluate the models on the BRIGHT benchmark, which requires complex multi-step reasoning and broad knowledge retrieval. Compared with the strongest RL baseline (ReasonRank-7B), \modelname-8B achieves a superior average NDCG@10 of 29.7. This performance is especially evident in knowledge-intensive subsets such as TheoT., TheoQ., and LeetCode, demonstrating a remarkable advantage in these highly challenging scenarios. While the models without tool invocation (e.g., RankGPT-4, Rank-R1-14B) severely hallucinate when their parametric knowledge is exhausted, \modelname~effectively avoids this by dynamically retrieving external evidence when uncertain. This validates that our reward design successfully helps the model overcome parametric limitations in reasoning-heavy tasks.

\begin{table*}[t]
\centering
\caption{Evaluation results (NDCG@10) across the reasoning-intensive BRIGHT benchmark datasets.}
\label{tab:bright_detailed_evaluation}
\resizebox{\textwidth}{!}{
\begin{tabular}{l l | c c c c c c c c c c c c | c }
\toprule
\multirow{2}{*}{\textbf{Model}} & \multirow{2}{*}{\textbf{Training}} & \multicolumn{13}{c}{\textbf{BRIGHT}} \\
\cmidrule(lr){3-15}
& & Bio. & Earth. & Econ. & Psy. & Rob. & Stack. & Sus. & Leet. & Pony & AoPS & TheoQ. & TheoT. & Avg \\
\midrule
BM25 & - & 18.2 & 27.9 & 16.4 & 13.4 & 10.9 & 16.3 & 16.1 & 4.3 & 24.7 & 6.5 & 2.1 & 7.3 & 13.7 \\
% \midrule
RankGPT-4 & zero-shot & 33.8 & 34.2 & 16.7 & 27.0 & 22.3 & 27.7 & 11.1 & 3.4 & 15.6 & 1.2 & 0.2 & 8.6 & 17.0
\\
\midrule
Rank1-7B & SFT & 31.4 & 36.7 & 18.3 & 25.4 & 13.8 & 17.6 & 24.8 & 16.7 & 9.5 & 6.1 & 9.5 & 11.6 & 18.5 \\
Rank1-14B & SFT & 29.6 & 34.8 & 17.2 & 24.3 & 18.6 & 16.2 & 24.5 & 17.5 & 14.4 & 5.5 & 9.2 & 10.7 & 18.5 \\
Reason-Point-4B & SFT & 23.6 & 29.0 & 15.0 & 23.7 & 16.7 & 12.2 & 18.3 & 18.4 & 12.4 & 8.9 & 11.0 & 9.4 & 16.5 \\
Direct-Point-4B & SFT & 34.9 & 45.1 & 23.3 & 31.8 & 26.6 & 23.6 & 30.7 & 18.5 & 35.4 & 7.2 & 13.6 & 15.2 & 25.5 \\
Reason-Point-8B & SFT & 24.9 & 34.6 & 17.5 & 26.2 & 25.9 & 22.4 & 19.7 & 11.9 & 36.6 & 9.3 & 6.5 & 12.6 & 20.7 \\
Direct-Point-8B & SFT & 33.9 & 46.4 & 24.6 & 31.6 & 25.8 & 25.9 & 32.0 & 25.3 & 35.5 & 12.0 & 13.5 & 15.2 & 26.8 \\
Reason-List-4B & SFT & 30.7 & 37.3 & 18.7 & 27.7 & 27.9 & 19.8 & 28.5 & 28.1 & 13.7 & 9.1 & 13.9 & 13.3 & 22.4 \\
Direct-List-4B & SFT & 32.7 & 38.6 & 20.0 & 28.4 & 28.6 & 20.5 & 31.2 & 30.9 & 15.1 & 10.4 & 17.8 & 15.6 & 24.1 \\
Reason-List-8B & SFT & 31.9 & 39.6 & 22.4 & 29.0 & 29.9 & 23.4 & 34.5 & 26.8 & 18.9 & 9.7 & 15.6 & 12.1 & 24.5 \\
Direct-List-8B & SFT & 32.6 & 38.4 & 21.3 & 28.9 & 31.9 & 22.6 & 31.8 & 28.9 & 16.9 & 11.1 & 18.5 & 15.4 & 24.9 \\
\midrule
TFRank-4B & SFT+GRPO & 33.2 & 45.9 & 17.6 & 29.5 & 21.0 & 20.9 & 18.3 & 25.0 & 9.1 & 9.5 & 9.8 & 7.3 & 20.6 \\
TFRank-8B & SFT+GRPO & 33.7 & 46.2 & 23.7 & 26.0 & 24.1 & 20.1 & 23.6 & 28.8 & 12.5 & 10.8 & 11.4 & 9.7 & 22.6 \\
ReasonRank-7B & SFT+GRPO & 36.3 & 44.2 & 24.8 & 31.7 & 30.7 & 24.9 & 32.8 & 28.7 & 17.5 & 12.0 & 18.5 & 14.0 & 26.4 \\
\midrule
REARANK-7B & GRPO & 23.4 & 27.4 & 18.5 & 24.2 & 17.4 & 16.3 & 25.1 & 27.0 & 8.0 & 7.4 & 7.9 & 9.5 & 17.7 \\
Rank-R1-7B & GRPO & 26.0 & 28.5 & 17.2 & 24.2 & 19.1 & 10.4 & 24.2 & 4.3 & 19.8 & 4.3 & 10.9 & 8.3 & 16.4 \\
Rank-R1-14B & GRPO & 31.2 & 38.5 & 21.2 & 26.4 & 22.6 & 18.9 & 27.5 & 20.2 & 9.2 & 9.7 & 9.2 & 11.9 & 20.5 \\
\midrule
\rowcolor{gray!15} \textbf{TALReranker}-4B & {SFT+GRPO} &  33.5 & 45.3 & 25.4 & 34.7 & 24.8 & 24.8 & 31.3 & 30.5 & 25.3 & 8.2 & 26.5 & 27.8 & \underline{28.2} \\
\rowcolor{gray!15} \textbf{TALReranker}-8B & {SFT+GRPO} &  37.1 & 43.0 & 25.3 & 37.8 & 26.3 & 23.4 & 31.2 & 30.7 & 38.3 & 8.0 & 26.5 & 28.9 & \textbf{29.7 }\\
\bottomrule
\end{tabular}
}
\end{table*}

\subsection{Efficiency of \modelname}

\textbf{Throughput Superiority against Baselines.} A major challenge for reasoning-based retrieval models is the severe latency caused by long text generation during inference. As shown in Figure~\ref{fig:efficiency}, \modelname~outperforms the traditional speed-accuracy tradeoff on the BRIGHT benchmark. Models relying on rigid internal reasoning suffer from severe latency; for example, Rank-R1-14B processes a mere 30 queries per hour. Conversely, direct-scoring models (e.g., Direct-Point, TFRank) are fast (500--1000 queries/hour) but lack the deep reasoning required for high accuracy. \modelname-4B and 8B achieve the best of both: they match the high throughput of direct-scoring models (977 and 778 queries/hour, respectively) while vastly outperforming them in precision. This proves that empowering a compact model with adaptive tool invocation is fundamentally more efficient than blindly scaling parameters or enforcing lengthy generation.

\begin{figure}[!t]
    \centering
    \includegraphics[width=\columnwidth]{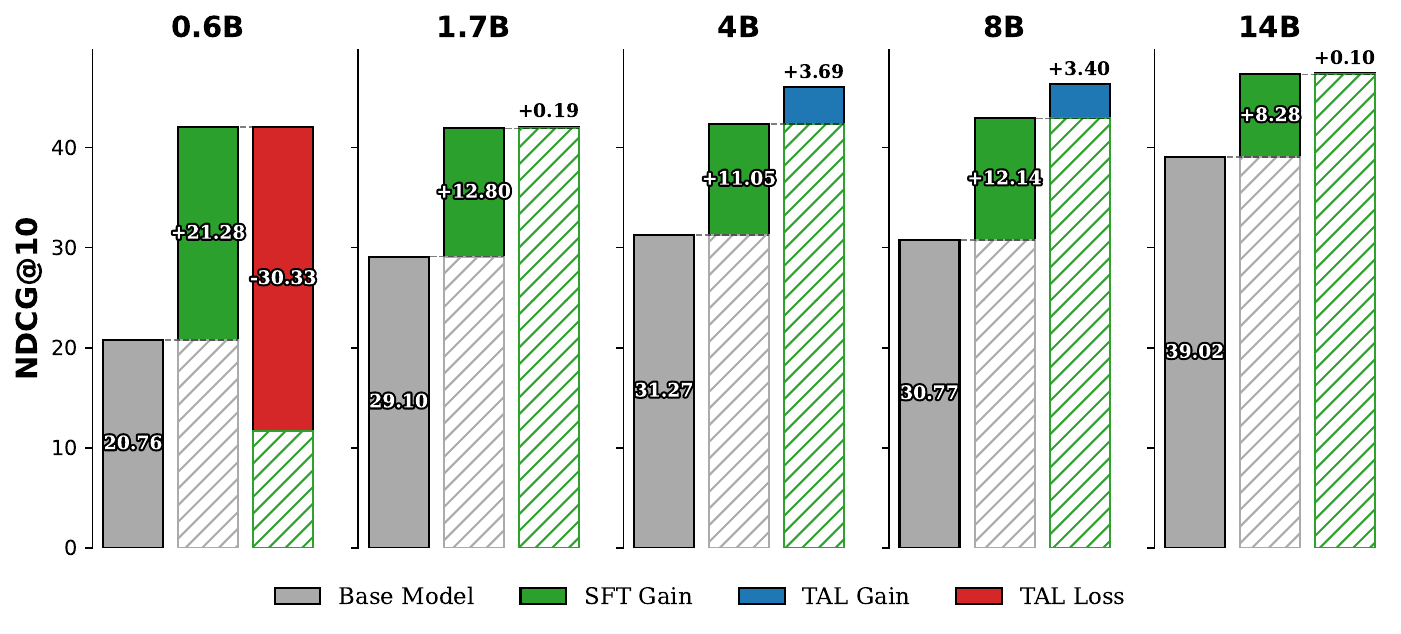}
    \caption{Ablation of training stages across model scales on the BEIR benchmark.}
    \label{fig:ablation_waterfall}
\end{figure}

\textbf{Adaptive Tool Invocation.} To validate our core claim of adaptive routing, we conduct a fine-grained statistical analysis of tool invocation rates across tasks of varying difficulty. 
Table~\ref{tab:adaptive_hops} demonstrates that our cost-aware RL successfully calibrates tool usage to the model's internal capacity.
Crucially, we observe a stark contrast in behavior between the simple, in-domain TREC-DL tasks and the complex, reasoning-intensive BRIGHT subsets. For instance, the 4B variant bypasses tools entirely on the simpler DL19 benchmark (0.00\% invocation rate), effectively acting as a fast parametric scorer. However, when confronted with OOD and knowledge-heavy tasks, it intelligently requests external evidence for complex edge cases (e.g., soaring to a 1.25\% invocation rate on Psychology and 1.47\% on Pony for the 1.7B model). 
Furthermore, this routing behavior scales logically with parameter size. The 0.6B model, lacking sufficient parametric knowledge, acts as a pure routing agent, relying on the search tool for 100\% of queries across all datasets, but the tool-calling format failed. Conversely, the robust 8B model rarely requires tools, reducing invocation on Psychology to a mere 0.28\%. By dynamically allocating computational budget strictly to uncertain queries, \modelname~provides concrete proof of adaptive epistemic boundary management, ensuring maximum efficiency without compromising accuracy.

\subsection{Ablation Study}

\begin{figure}[t]
    \centering
    \includegraphics[width=0.9\columnwidth]{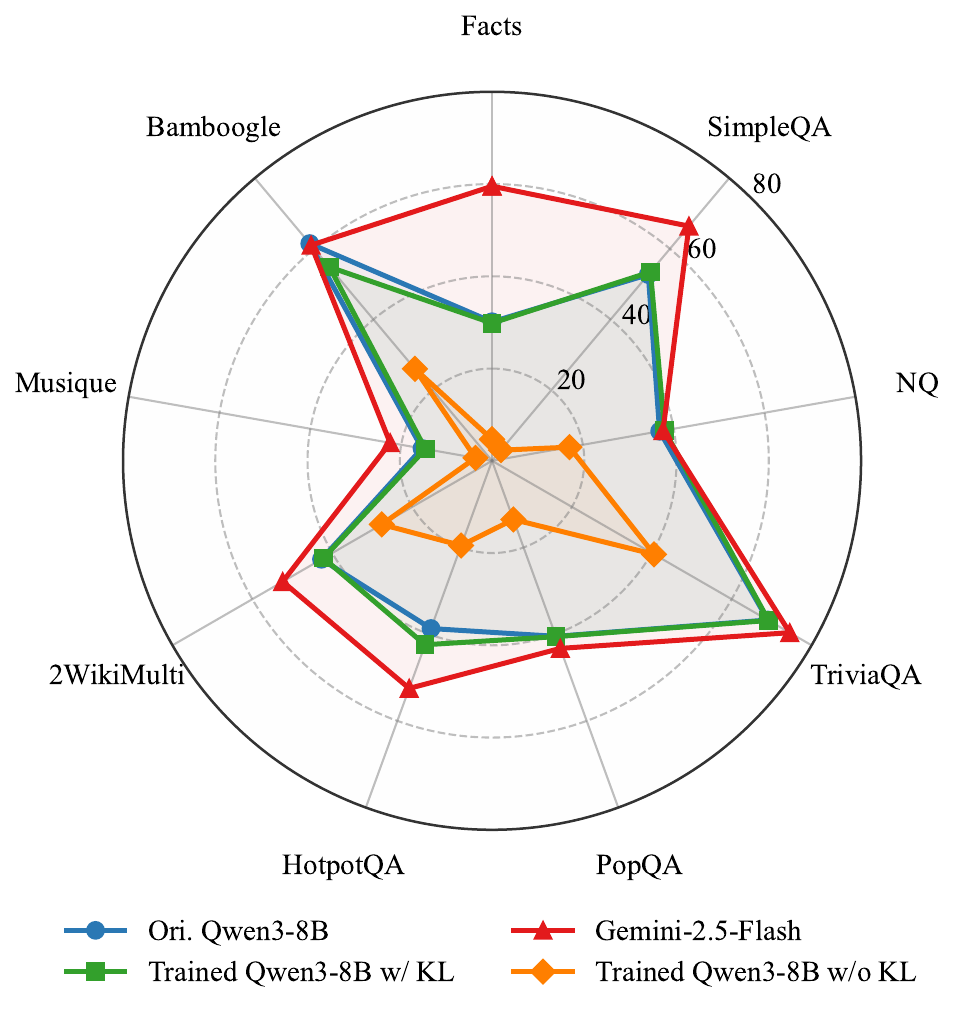}
    \caption{Generative performance on nine QA benchmarks. }
    \label{fig:radar}
\end{figure}
\textbf{Two-Stage Training Ablation.} To dissect the contribution of each training phase, Figure~\ref{fig:ablation_waterfall} visualizes the performance impact of our two-stage training paradigm across model scales. While pointwise SFT consistently improves performance across all sizes, directly applying RL causes severe degradation in smaller models. Specifically, the 0.6B model loses $30.33$ absolute NDCG@10 points. Error analysis reveals its inability to handle the complex action space, frequently missing the terminal \texttt{<answer>} token and resulting in zero-reward episodes. As capacity marginally increases to 1.7B, the model avoids structural collapse.
By safely anchoring the language-preserved scoring, the tool-adaptive learning successfully unlocks significant gains in reasoning for the 4B and 8B variants, enabling the reranker to make accurate judgments based on external knowledge.

\textbf{The Necessity of KL Regularization.} To isolate the impact of the language-preserving loss, we present a component experiment to demonstrate the importance of this constraint. Specifically, we conduct agentic search experiments using common benchmarks to measure the ability of a Qwen3-8B before and after training, including nine sub-benchmarks: NQ~\cite{kwiatkowski-etal-2019-natural}, TriviaQA~\cite{joshi-etal-2017-triviaqa}, PopQA~\cite{mallen-etal-2023-trust}, HotpotQA~\cite{yang-etal-2018-hotpotqa}, 2WikiMultiHopQA~\cite{ho-etal-2020-constructing}, Musique~\cite{trivedi2022musiquemultihopquestionssinglehop}, Bamboogle~\cite{press-etal-2023-measuring}, Facts~\cite{cheng2025factsleaderboardcomprehensivebenchmark}, and SimpleQA~\cite{wei2024measuringshortformfactualitylarge}. The exact matching answer results are in Figure~\ref{fig:radar}. 
As observed, training reranker exclusively with the pointwise objective (Trained Qwen3-8B w/o KL) reduces the model to a rigid scalar function. Deprived of the KL penalty, the policy over-optimizes relevance regression at the expense of its native linguistic and factual capacities.
This leads to severe catastrophic forgetting and a systemic collapse across all generative QA tasks (e.g., performance on SimpleQA and Facts drops precipitously to near zero).
In stark contrast, incorporating the masked KL constraint (Trained Qwen3-8B w/ KL) strictly anchors the policy to the base model (Ori. Qwen3-8B) during sequence generation. This effectively preserves the original parametric knowledge, ensuring a robust and factual foundation for subsequent tool-augmented explorations.

\textbf{Sensitivity to Reward Hyperparameters.} Our cost-aware reward fundamentally relies on the penalty intensity $\lambda$ and the threshold $\gamma$. As detailed in the Appendix, empirical ablations demonstrate that the model behavior is highly sensitive to these bounds. A strict penalty ($\lambda \ge 0.1$) rapidly suppresses tool invocation but forces blind guessing, degrading overall NDCG. Conversely, the threshold $\gamma$ dictates the epistemic confidence boundary; a relaxed threshold ($\gamma \ge 0.5$) induces overconfidence and severe hallucination penalties. We find that a moderate penalty ($\lambda=0.02$) paired with an optimal threshold ($\gamma \in [0.2, 0.3]$) perfectly balances internal reliance against the necessity for external validation, yielding peak accuracy without unnecessary search latency.

\subsection{Case Study}

To intuitively illustrate how \modelname~dynamically manages its epistemic boundaries, we provide detailed inference trajectories in the Appendix. The case studies show the tool-adaptive routing mechanism. In scenarios involving factual conflicts or high uncertainty (e.g., verifying absolute planetary temperatures to validate a flawed premise), the model correctly suspends generation, formulates a \texttt{<tool\_call>}, and integrates environmental feedback before final scoring. Conversely, when evaluating fundamental concepts well within its parametric memory, the model confidently relies on its internal knowledge, emitting a direct \texttt{<answer>} without incurring tool-use latency. This contrast highlights the effectiveness of the cost-aware reward in preventing both over-reliance on external tools and hallucinatory guesses.

\section{Conclusion}

In this paper, we introduced \modelname, a novel tool-adaptive framework that fundamentally bridges the gap between static discriminative reranking and dynamic generative reasoning. By modeling the pointwise scoring process as an agentic MDP, we successfully integrated a cost-aware RL strategy equipped with an asymmetric penalty and a language-preserving hybrid loss. This design enables the model to autonomously manage its epistemic boundaries, bypassing tools to maintain high efficiency on standard tasks while intelligently requesting external evidence to prevent hallucinations on complex reasoning benchmarks like BRIGHT. While \modelname~establishes a new state-of-the-art across both in-domain and out-of-domain retrieval tasks, a notable limitation is its strict dependency on the base LLM's intrinsic reasoning capabilities, especially concerning the tool-use trajectories. Future work will explore targeted distillation techniques and robust constraints to decouple these adaptive behaviors from internal reasoning limits, thereby enhancing the stability of the framework.

\newpage
\bibliography{custom}

@inproceedings{qu2025from,
title={From Exploration to Mastery: Enabling {LLM}s to Master Tools via Self-Driven Interactions},
author={Changle Qu and Sunhao Dai and Xiaochi Wei and Hengyi Cai and Shuaiqiang Wang and Dawei Yin and Jun Xu and Ji-Rong Wen},
pages = {1--23},
booktitle={International Conference on Learning Representations},
year={2025},
}

@inproceedings{erdogan-etal-2024-tinyagent,
    title = {{T}iny{A}gent: Function Calling at the Edge},
    author = {Erdogan, Lutfi Eren  and
      Lee, Nicholas  and
      Jha, Siddharth  and
      Kim, Sehoon  and
      Tabrizi, Ryan  and
      Moon, Suhong  and
      Hooper, Coleman Richard Charles  and
      Anumanchipalli, Gopala  and
      Keutzer, Kurt  and
      Gholami, Amir},
    booktitle = {Empirical Methods in Natural Language Processing},
    pages = {80--88},
    year={2024}
}

@inproceedings{
schick2023toolformer,
title={Toolformer: Language Models Can Teach Themselves to Use Tools},
author={Timo Schick and Jane Dwivedi-Yu and Roberto Dessi and Roberta Raileanu and Maria Lomeli and Eric Hambro and Luke Zettlemoyer and Nicola Cancedda and Thomas Scialom},
booktitle={Neural Information Processing System},
pages = {68539 - 68551},
year={2023}
}

@inproceedings{yao2023reactsynergizingreasoningacting,
title={ReAct: Synergizing Reasoning and Acting in Language Models}, 
author={Shunyu Yao and Jeffrey Zhao and Dian Yu and Nan Du and Izhak Shafran and Karthik Narasimhan and Yuan Cao},
pages = {1 - 33},
booktitle={International Conference on Learning Representations},
year={2023}
}

@article{yang2026efficientagentsmemorytool,
title={Toward Efficient Agents: Memory, Tool learning, and Planning}, 
author={Xiaofang Yang and Lijun Li and Heng Zhou and Tong Zhu and Xiaoye Qu and Yuchen Fan and Qianshan Wei and Rui Ye and Li Kang and Yiran Qin and Zhiqiang Kou and Daizong Liu and Qi Li and Ning Ding and Siheng Chen and Jing Shao},
journal={arXiv preprint arXiv:2601.14192},
year={2026}
}

@inproceedings{gao-etal-2025-efficient,
  title={Efficient Tool Use with Chain-of-Abstraction Reasoning},
  author={Gao, Silin  and
      Dwivedi-Yu, Jane  and
      Yu, Ping  and
      Tan, Xiaoqing Ellen  and
      Pasunuru, Ramakanth  and
      Golovneva, Olga  and
      Sinha, Koustuv  and
      Celikyilmaz, Asli  and
      Bosselut, Antoine  and
      Wang, Tianlu},
  booktitle={International Conference on Learning Representations},
  pages={2727--2743},
  year={2025}
}

@article{su2025toolorchestraelevatingintelligenceefficient,
  title={ToolOrchestra: Elevating Intelligence via Efficient Model and Tool Orchestration},
  author={Hongjin Su and Shizhe Diao and Ximing Lu and Mingjie Liu and Jiacheng Xu and Xin Dong and Yonggan Fu and Peter Belcak and Hanrong Ye and Hongxu Yin and Yi Dong and Evelina Bakhturina and Tao Yu and Yejin Choi and Jan Kautz and Pavlo Molchanov},
  journal={arXiv preprint arXiv:2511.21689},
  year={2025}
}

@inproceedings{qian-etal-2025-smart,
    title = {{SMART}: Self-Aware Agent for Tool Overuse Mitigation},
    author = {Qian, Cheng  and
      Acikgoz, Emre Can  and
      Wang, Hongru  and
      Chen, Xiusi  and
      Sil, Avirup  and
      Hakkani-T{\"u}r, Dilek  and
      Tur, Gokhan  and
      Ji, Heng},
    booktitle = {Findings of the Association for Computational Linguistics},
    pages = {4604--4621},
    year = {2025}
}

@inproceedings{
yoon2025acurank,
title={AcuRank: Uncertainty-Aware Adaptive Computation for Listwise Reranking},
author={Soyoung Yoon and Gyuwan Kim and GYU-HWUNG CHO and seung-won hwang},
booktitle={Neural Information Processing System},
pages = {26388--26417},
year={2025}
}

@article{zhuang2025rankr1enhancingreasoningllmbased,
  author={Shengyao Zhuang and Xueguang Ma and Bevan Koopman and Jimmy Lin and Guido Zuccon},
  title={Rank-R1: Enhancing Reasoning in LLM-based Document Rerankers via Reinforcement Learning},
  journal={arXiv preprint arXiv:2503.06034},
  year={2025}
}

@inproceedings{zhang2025rearankreasoningrerankingagent,
    title = {REARANK: Reasoning Re-ranking Agent via Reinforcement Learning},
    author = {Zhang, Le  and
      Wang, Bo  and
      Qiu, Xipeng  and
      Reddy, Siva  and
      Agrawal, Aishwarya},
    booktitle = {Empirical Methods in Natural Language Processing},
    pages = {2458--2471},
    year = {2025}
}

@inproceedings{fan2025tfrankthinkfreereasoningenables,
      title={TFRank: Think-Free Reasoning Enables Practical Pointwise LLM Ranking}, 
      author={Yongqi Fan and Xiaoyang Chen and Dezhi Ye and Jie Liu and Haijin Liang and Jin Ma and Ben He and Yingfei Sun and Tong Ruan},
        booktitle={AAAI Conference on Artificial Intelligence},
      page={21020-21028},
      year={2025}
}

@article{craswell2020overviewtrec2019deep,
      title={Overview of the TREC 2019 deep learning track}, 
      author={Nick Craswell and Bhaskar Mitra and Emine Yilmaz and Daniel Campos and Ellen M. Voorhees},
      journal={arXiv preprint arXiv:2003.07820},
      year={2020}
}

@article{craswell2021overviewtrec2020deep,
      title={Overview of the TREC 2020 deep learning track}, 
      author={Nick Craswell and Bhaskar Mitra and Emine Yilmaz and Daniel Campos},
      journal={arXiv preprint arXiv:2102.07662},
      year={2021}
}

@article{yang2025qwen3,
  title={Qwen3 technical report},
  author={Yang, An and Li, Anfeng and Yang, Baosong and others},
  journal={arXiv preprint arXiv:2505.09388},
  year={2025}
}

@inproceedings{
su2025bright,
title={{BRIGHT}: A Realistic and Challenging Benchmark for Reasoning-Intensive Retrieval},
author={Hongjin SU and Howard Yen and Mengzhou Xia and Weijia Shi and Niklas Muennighoff and Han-yu Wang and Liu Haisu and Quan Shi and Zachary S Siegel and Michael Tang and Ruoxi Sun and Jinsung Yoon and Sercan O Arik and Danqi Chen and Tao Yu},
  pages={1-51},
booktitle={International Conference on Learning Representations},
year={2025}
}

@inproceedings{
thakur2021beir,
title={{BEIR}: A Heterogeneous Benchmark for Zero-shot Evaluation of Information Retrieval Models},
author={Nandan Thakur and Nils Reimers and Andreas R{\"u}ckl{\'e} and Abhishek Srivastava and Iryna Gurevych},
booktitle={Neural Information Processing System},
year={2021}
}

@inproceedings{10.1145/3404835.3463238, 
  author={Jimmy Lin and Xueguang Ma and Sheng-Chieh Lin and Jheng-Hong Yang and Ronak Pradeep and Rodrigo Frassetto Nogueira},
  title={Pyserini: A Python Toolkit for Reproducible Information Retrieval Research with Sparse and Dense Representations},
  booktitle={International ACM SIGIR Conference},
  pages={2356-2362},
  year={2021}
}

@inproceedings{sun2023is,
title={Is Chat{GPT} Good at Search? Investigating Large Language Models as Re-Ranking Agents},
author={Weiwei Sun and Lingyong Yan and Xinyu Ma and Shuaiqiang Wang and Pengjie Ren and Zhumin Chen and Dawei Yin and Zhaochun Ren},
booktitle={Empirical Methods in Natural Language Processing},
page={14918--14937},
year={2023}
}

@inproceedings{
weller2025rank,
title={Rank1: Test-Time Compute for Reranking in Information Retrieval},
author={Orion Weller and Kathryn Ricci and Eugene Yang and Andrew Yates and Dawn Lawrie and Benjamin Van Durme},
pages = {1–18}, 
booktitle={Conference on Language Modeling},
year={2025}
}

@inproceedings{10.1145/3770854.3783917, 
author = {Zeng, Ziyang and Jing, Heming and Chen, Jindong and Li, Xiangli and Liu, Hongyu and He, Yixuan and Li, Zhengyu and Sun, Yige and Xie, Zheyong and Yang, Yuqing and Cao, Shaosheng and Fan, Jun and Wu, Yi and Hu, Yao}, 
title = {Optimizing Generative Ranking Relevance via Reinforcement Learning in Xiaohongshu Search}, 
booktitle={ACM SIGKDD Conference},
pages = {2551–2561}, 
year={2026}
}

@inproceedings{wang2024rcagent,
  title={Rcagent: Cloud Root Cause Analysis by Autonomous Agents with Tool-augmented Large Language Models},
  author={Wang, Zefan and Liu, Zichuan and Zhang, Yingying and Zhong, Aoxiao and Wang, Jihong and Yin, Fengbin and Fan, Lunting and Wu, Lingfei and Wen, Qingsong},
  booktitle={International Conference on Information and Knowledge Management},
  pages={4966--4974},
  year={2024}
}

@inproceedings{xu2026beyond,
title={Beyond Sequential Reranking: Reranker-Guided Search Improves Reasoning Intensive Retrieval},
author={Haike Xu and Tong Chen},
  pages={1--15},
booktitle={International Conference on Learning Representations},
year={2026}
}

@article{
liang2023holistic,
title={Holistic Evaluation of Language Models},
author={Percy Liang and Rishi Bommasani and Tony Lee and others},
journal={Transactions on Machine Learning Research},
page={2835-8856},
year={2023}
}

@inproceedings{li2026prorankpromptwarmupreinforcement,
    title = {ProRank: Prompt Warmup via Reinforcement Learning for Small Language Models Reranking},
    author = {LI, Xianming  and
      Shakir, Aamir  and
      Huang, Rui  and
      Lipp, Julius  and
      Clavi{\'e}, Benjamin  and
      Li, Jing},
    booktitle = {Association for Computational Linguistics},
    pages = {1026--1037},
    year = {2026}
}

@inproceedings{qin-etal-2024-large,
    title = {Large Language Models are Effective Text Rankers with Pairwise Ranking Prompting},
    author = {Qin, Zhen  and
      Jagerman, Rolf  and
      Hui, Kai  and
      Zhuang, Honglei  and
      Wu, Junru  and
      Yan, Le  and
      Shen, Jiaming  and
      Liu, Tianqi  and
      Liu, Jialu  and
      Metzler, Donald  and
      Wang, Xuanhui  and
      Bendersky, Michael},
  booktitle={Findings of the Association for Computational Linguistics},
    pages = {1504--1518},
    year={2024}
}

@inproceedings{luo-etal-2024-prp,
    title = {{PRP}-Graph: Pairwise Ranking Prompting to {LLM}s with Graph Aggregation for Effective Text Re-ranking},
    author = {Luo, Jian  and
      Chen, Xuanang  and
      He, Ben  and
      Sun, Le},
    booktitle = {Association for Computational Linguistics},
    year = {2024},
    pages = {5766--5776}
}

@article{li2026semanticsimilarityrethinkingretrieval,
      title={Beyond Semantic Similarity: Rethinking Retrieval for Agentic Search via Direct Corpus Interaction}, 
      author={Zhuofeng Li and Haoxiang Zhang and Cong Wei and Pan Lu and Ping Nie and Yi Lu and Yuyang Bai and Shangbin Feng and Hangxiao Zhu and Ming Zhong and Yuyu Zhang and Jianwen Xie and Yejin Choi and James Zou and Jiawei Han and Wenhu Chen and Jimmy Lin and Dongfu Jiang and Yu Zhang},
      journal={arXiv preprint arXiv:2605.05242},
      year={2026}
}

@inproceedings{weller-etal-2025-followir,
    title = {{F}ollow{IR}: Evaluating and Teaching Information Retrieval Models to Follow Instructions},
    author = {Weller, Orion  and
      Chang, Benjamin  and
      MacAvaney, Sean  and
      Lo, Kyle  and
      Cohan, Arman  and
      Van Durme, Benjamin  and
      Lawrie, Dawn  and
      Soldaini, Luca},
  booktitle={Nations of the Americas Chapter of the Association for Computational Linguistics},
    year = {2025},
    pages = {11926--11942}
}

@inproceedings{mao-etal-2024-chatretriever,
    title = {{C}hat{R}etriever: Adapting Large Language Models for Generalized and Robust Conversational Dense Retrieval},
    author = {Mao, Kelong  and
      Deng, Chenlong  and
      Chen, Haonan  and
      Mo, Fengran  and
      Liu, Zheng  and
      Sakai, Tetsuya  and
      Dou, Zhicheng},
    booktitle = {Empirical Methods in Natural Language Processing},
    year = {2024},
    pages = {1227--1240}
}

@article{liu2025sample,
  title={Sample-efficient LLM Optimization with Reset Replay},
  author={Liu, Zichuan and Wang, Jinyu and Song, Lei and Bian, Jiang},
  journal={arXiv preprint arXiv:2508.06412},
  year={2025}
}

@article{Guo_2025,
   title={DeepSeek-R1 incentivizes reasoning in LLMs through reinforcement learning},
   journal={Nature},
   author={Guo, Daya and Yang, Dejian and Zhang, Haowei and others},
   year={2025}, 
   pages={633–638} 
}

@inproceedings{
yu2024rankrag,
title={Rank{RAG}: Unifying Context Ranking with Retrieval-Augmented Generation in {LLM}s},
author={Yue Yu and Wei Ping and Zihan Liu and Boxin Wang and Jiaxuan You and Chao Zhang and Mohammad Shoeybi and Bryan Catanzaro},
booktitle={Neural Information Processing System},
page={121156 - 121184},
year={2024}
}

@inproceedings{
wei2022chain,
title={Chain of Thought Prompting Elicits Reasoning in Large Language Models},
author={Jason Wei and Xuezhi Wang and Dale Schuurmans and Maarten Bosma and brian ichter and Fei Xia and Ed H. Chi and Quoc V Le and Denny Zhou},
booktitle={Neural Information Processing System},
page={24824 - 24837},
year={2022}
}

@article{li2025matchinggenerationsurveygenerative,
      title={From Matching to Generation: A Survey on Generative Information Retrieval}, 
      author={Xiaoxi Li and Jiajie Jin and Yujia Zhou and Yuyao Zhang and Peitian Zhang and Yutao Zhu and Zhicheng Dou},
      journal={Transactions on Information Systems},
      page={1 - 62},
      year={2025}
}

@article{singh2025agentic,
      title={Agentic Reasoning and Tool Integration for {LLM}s via Reinforcement Learning}, 
      author={Joykirat Singh and Yash Pandya and Pranav Vajreshwari and Raghav Magazine and Akshay Nambi},
      journal={arXiv preprint arXiv:2505.01441},
      year={2025}
      }

@inproceedings{
lu2026rethinking,
title={Rethinking Reasoning in Document Ranking: Why Chain-of-Thought Falls Short},
author={Xuan Lu and Haohang Huang and Rui Meng and Yaohui Jin and Wenjun Zeng and Xiaoyu Shen},
      page={1 - 20},
booktitle={International Conference on Learning Representations},
year={2026}
}

@inproceedings{
chern2025factool,
title={FacTool: Factuality Detection in Generative {AI} -- A Tool Augmented Framework for Multi-Task and Multi-Domain Scenarios},
author={Ethan Chern and Steffi Chern and Shiqi Chen and Weizhe Yuan and Kehua Feng and Chunting Zhou and Junxian He and Graham Neubig and Pengfei Liu},
      page={1 - 28},
booktitle={Conference on Language Modeling},
year={2025}
}

@inproceedings{
gou2024critic,
title={{CRITIC}: Large Language Models Can Self-Correct with Tool-Interactive Critiquing},
author={Zhibin Gou and Zhihong Shao and Yeyun Gong and yelong shen and Yujiu Yang and Nan Duan and Weizhu Chen},
      page={1 - 78},
booktitle={International Conference on Learning Representations},
year={2024}
}

@inproceedings{Zhuang_2024,
   title={A Setwise Approach for Effective and Highly Efficient Zero-shot Ranking with Large Language Models},
   author={Zhuang, Shengyao and Zhuang, Honglei and Koopman, Bevan and Zuccon, Guido},
  booktitle={International ACM SIGIR Conference},
   page={38-47},
   year={2024}
}

@article{liu2026reasonrankempoweringpassageranking,
      title={ReasonRank: Empowering Passage Ranking with Strong Reasoning Ability}, 
      author={Wenhan Liu and Xinyu Ma and Weiwei Sun and Yutao Zhu and Yuchen Li and Dawei Yin and Zhicheng Dou},
      journal={arXiv preprint arXiv:2508.07050},
      year={2026}
}

@article{kwiatkowski-etal-2019-natural,
      title={Natural Questions: A Benchmark for Question Answering Research}, 
      author={Kwiatkowski, Tom  and
      Palomaki, Jennimaria  and
      Redfield, Olivia  and
      Collins, Michael  and
      Parikh, Ankur  and
      Alberti, Chris  and
      Epstein, Danielle  and
      Polosukhin, Illia  and
      Devlin, Jacob  and
      Lee, Kenton  and
      Toutanova, Kristina  and
      Jones, Llion  and
      Kelcey, Matthew  and
      Chang, Ming-Wei  and
      Dai, Andrew M.  and
      Uszkoreit, Jakob  and
      Le, Quoc  and
      Petrov, Slav},
      journal={Transactions of the Association for Computational Linguistics},
      page={452--466},
      year={2019}
}

@inproceedings{joshi-etal-2017-triviaqa,
    title = {{T}rivia{QA}: A Large Scale Distantly Supervised Challenge Dataset for Reading Comprehension},
    author = {Joshi, Mandar  and
      Choi, Eunsol  and
      Weld, Daniel  and
      Zettlemoyer, Luke},
    booktitle = {Association for Computational Linguistics},
    pages = {1601--1611},
    year = {2017}
}

@inproceedings{mallen-etal-2023-trust,
    title = {When Not to Trust Language Models: Investigating Effectiveness of Parametric and Non-Parametric Memories},
    author = {Mallen, Alex  and
      Asai, Akari  and
      Zhong, Victor  and
      Das, Rajarshi  and
      Khashabi, Daniel  and
      Hajishirzi, Hannaneh},
    booktitle = {Association for Computational Linguistics},
    pages = {9802--9822},
    year = {2023}
}

@inproceedings{yang-etal-2018-hotpotqa,
    title = {{H}otpot{QA}: A Dataset for Diverse, Explainable Multi-hop Question Answering},
    author = {Yang, Zhilin  and
      Qi, Peng  and
      Zhang, Saizheng  and
      Bengio, Yoshua  and
      Cohen, William  and
      Salakhutdinov, Ruslan  and
      Manning, Christopher D.},
    booktitle = {Empirical Methods in Natural Language Processing},
    pages = {2369--2380},
    year = {2018}
}

@inproceedings{ho-etal-2020-constructing,
    title = {Constructing A Multi-hop {QA} Dataset for Comprehensive Evaluation of Reasoning Steps},
    author = {Ho, Xanh  and
      Duong Nguyen, Anh-Khoa  and
      Sugawara, Saku  and
      Aizawa, Akiko},
    booktitle = {International Committee on Computational Linguistics},
    pages = {6609--6625},
    year = {2020}
}

@article{trivedi2022musiquemultihopquestionssinglehop,
  title={MuSiQue: Multihop Questions via Single-hop Question Composition},
  author={Harsh Trivedi and Niranjan Balasubramanian and Tushar Khot and Ashish Sabharwal},
  journal={arXiv preprint arXiv:2108.00573},
  year={2022}
}

@inproceedings{press-etal-2023-measuring,
    title = {Measuring and Narrowing the Compositionality Gap in Language Models},
    author = {Press, Ofir  and
      Zhang, Muru  and
      Min, Sewon  and
      Schmidt, Ludwig  and
      Smith, Noah  and
      Lewis, Mike},
    booktitle = {Findings of the Empirical Methods in Natural Language Processing},
    pages = {5687--5711},
    year = {2023}
}

@article{wei2024measuringshortformfactualitylarge,
  title={Measuring short-form factuality in large language models},
  author={Jason Wei and Nguyen Karina and Hyung Won Chung and Yunxin Joy Jiao and Spencer Papay and Amelia Glaese and John Schulman and William Fedus},
  journal={arXiv preprint arXiv:2411.04368},
  year={2024}
}

@article{cheng2025factsleaderboardcomprehensivebenchmark,
  title={The FACTS Leaderboard: A Comprehensive Benchmark for Large Language Model Factuality},
  author={Aileen Cheng and Alon Jacovi and Amir Globerson and others},
  journal={arXiv preprint arXiv:2512.10791},
  year={2025}
}

@article{zhang2025qwen3embeddingadvancingtext,
  title={Qwen3 Embedding: Advancing Text Embedding and Reranking Through Foundation Models},
  author={Yanzhao Zhang and Mingxin Li and Dingkun Long and Xin Zhang and Huan Lin and Baosong Yang and Pengjun Xie and An Yang and Dayiheng Liu and Junyang Lin and Fei Huang and Jingren Zhou},
  journal={arXiv preprint arXiv:2506.05176},
  year={2025}
}

@article{li2026qwen3vlembeddingqwen3vlrerankerunifiedframework,
  title={Qwen3-VL-Embedding and Qwen3-VL-Reranker: A Unified Framework for State-of-the-Art Multimodal Retrieval and Ranking},
  author={Mingxin Li and Yanzhao Zhang and Dingkun Long and Keqin Chen and Sibo Song and Shuai Bai and Zhibo Yang and Pengjun Xie and An Yang and Dayiheng Liu and Jingren Zhou and Junyang Lin},
  journal={arXiv preprint arXiv:2601.04720},
  year={2026}
}

\newpage
\appendix
\section{Prompt Template}
\label{app:prompt}
The exact prompt templates used by \modelname{} during both training and inference are presented below. The prompts are divided into a system instruction and a dynamic user instruction that formats the target webpage features.

\begin{figure}[htbp]
\centering
\begin{tcolorbox}[
    colback=gray!4!white,
    colframe=gray!65!black,
    title=\textbf{System Prompt},
    fonttitle=\bfseries,
    fontupper=\small\sffamily,
    arc=1.5mm,
    boxrule=0.6pt,
    left=2mm, right=2mm, top=2mm, bottom=2mm
]
\textbf{\#\# Background information} \\
$\bullet$ Today is \texttt{\{cur\_date\}} \\
$\bullet$ You are Deep AI search and quality evaluator Assistant \\

Evaluate a webpage for a given query and answer whether the webpage is relevant based on how well the webpage satisfies the user's search intent. Note that the answer can only be yes or no for webpage. \\

I will provide you with many tools to help you evaluate the query, but hope you can minimize the use of tools as much as possible. Your output format should be one of the following two formats:

\vspace{0.5em}
\begin{tcolorbox}[colback=white, colframe=gray!40, arc=0mm, boxrule=0.5pt, left=1mm, top=1mm, bottom=1mm, right=1mm]
\texttt{<reasoning>}\\
\\
\texttt{</reasoning>} \\
\texttt{<answer>}\\
\texttt{yes|no}\\
\texttt{</answer>}
\end{tcolorbox}
\vspace{0.2em}
or
\vspace{0.2em}
\begin{tcolorbox}[colback=white, colframe=gray!40, arc=0mm, boxrule=0.5pt, left=1mm, top=1mm, bottom=1mm, right=1mm]
\texttt{<reasoning>} \\
\texttt{YOUR THINKING PROCESS} \\
\texttt{</reasoning>} \\
\texttt{<tool\_call>} \\
\texttt{JSON TOOL CALL} \\
\texttt{</tool\_call>}
\end{tcolorbox}
\vspace{0.5em}

You should always follow the above two formats strictly. \\
Only output the final answer inside the \texttt{<answer></answer>} tag, without any explanations or extra information.
\end{tcolorbox}
\end{figure}\vspace{-5mm}

\begin{figure}[!t]
\centering
\begin{tcolorbox}[
    colback=gray!4!white,
    colframe=gray!65!black,
    title=\textbf{User Prompt},
    fonttitle=\bfseries,
    fontupper=\small\sffamily,
    arc=1.5mm,
    boxrule=0.6pt,
    left=2mm, right=2mm, top=2mm, bottom=2mm
]
Evaluate a webpage is relevant to the given query, please judge whether the webpage satisfies the user's information need. You must conduct reasoning inside \texttt{<reasoning>} and \texttt{</reasoning>} first every time you get new information, THINKING PROCESS can be \texttt{<reasoning></reasoning>} If it's easy to determine. After reasoning, you can call tools in schemas by \texttt{<tool\_call> JSON\_CALL </tool\_call>}, and it will return the tool information between \texttt{<tool\_response>} and \texttt{</tool\_response>}. The information in \texttt{<tool\_response>} and \texttt{</tool\_response>} is a content of the tool you called, not the webpage should be rated. After \texttt{</tool\_response>} you should first \texttt{<reasoning>} again. You can call the tools many times as your want with semantically refining the user query or searching external knowledge. If you find no further external knowledge needed, you can directly provide the output inside \texttt{<answer>} and \texttt{</answer>}. Answer whether this webpage is satisfied, in the format of "yes" or "no"; yes=the webpage is satisfied user intent, no=not satisfied. For example, if you think a webpage is related to the query, output \texttt{<reasoning>\textbackslash{}n\textbackslash{}n</reasoning>}\texttt{<answer> yes</answer>} or \texttt{<reasoning>YOUR THINKING PROCESS</reasoning><tool\_call>JSON TOOL CALL</tool\_call>}. \\

\textbf{Now given:} \\
\textbf{Query:} \texttt{\{query\}} \\
\textbf{Webpage:} \texttt{\{docs\}}

\vspace{0.8em}
\hrule
\vspace{0.8em}

\textbf{[Webpage Feature Format (\texttt{\{docs\}})]} \\
\texttt{<Title>}: \texttt{\{title\}} \\
\texttt{<URL>}: \texttt{\{url\}} \\
\texttt{<Body>}: \texttt{\{body\}}
\end{tcolorbox}
\end{figure}

\section{Inference Execution Flow}
\label{app:inference}
During the inference stage, \modelname{} operates as an interactive agent, executing the exact Markov decision process defined in Methodology. The inference pipeline relies on intercepting specific tokens during autoregressive decoding to manage environmental transitions. Algorithm~\ref{alg:inference} formalizes this adaptive execution flow.

\begin{algorithm}[tb]
\caption{Inference Flow of TALRanker}
\label{alg:inference}
\textbf{Input}: User query $q$, candidate document $d$, generative policy $\pi_\theta$, maximum hops $K$, a search tool. \\
\textbf{Output}: Final score $s \in [0, 1]$.
\begin{algorithmic}[1] %[1] enables line numbers
\STATE Initialize context state: $h_0 \leftarrow \text{Prompt}(q, d)$.
\STATE Initialize external knowledge buffer: $\mathcal{E} \leftarrow \emptyset$.
\STATE Set hop counter: $t \leftarrow 0$.
\WHILE{$t \le K$}
    \STATE Autoregressively sample reasoning trajectory: $\tau \sim \pi_\theta(\cdot \mid h_t)$.
    \STATE Extract internal thinking process by \texttt{<reasoning>} from $\tau$.
    \STATE Extract terminal action $a_{\text{term}}$.
    \IF{$a_{\text{term}} == \texttt{<answer>}$}
        \STATE Extract probabilities for specific target tokens $P_{\text{yes}}$ and $P_{\text{no}}$ given $h_t \oplus \tau$.
        \STATE Calculate continuous score: $s \leftarrow \frac{P_{\text{yes}}}{P_{\text{yes}} + P_{\text{no}}}$.
        \RETURN $s$
    \ELSIF{$a_{\text{term}} == \texttt{<tool\_call>}$}
        \STATE Extract generated search query $q'$ from $\tau$.
        \STATE Execute external retrieval to fetch evidence: $e \leftarrow \text{Search}(q')$.
        \STATE $\mathcal{E} \leftarrow \mathcal{E} \cup \{e\}$.
        \STATE Update context state: $h_{t+1} \leftarrow h_t \oplus \tau \oplus \text{Format}(e)$.
        \STATE $t \leftarrow t + 1$.
    \ELSE
        \RETURN Fallback score (e.g., $0.0$).
    \ENDIF
\ENDWHILE
\STATE Calculate score $s$ directly from $h_{t}$ as in Eq.~\ref{eq:score}.
\RETURN $s$
\end{algorithmic}
\end{algorithm}

As detailed in Algorithm~\ref{alg:inference}, the decoding process is explicitly suspended upon generating a \texttt{<tool\_call>}$q'$\texttt{</tool\_call>} token. The system executes the external search API, concatenates the retrieved context into the state sequence, and resumes generation. This iterates until the \texttt{<answer>}(\texttt{yes}|\texttt{no})\texttt{</answer>} tag is emitted. To support fine-grained document sorting, the system extracts the exact log-probabilities of the \texttt{yes} and \texttt{no} tokens at the terminal step, computing the normalized continuous score $s$ to establish a strict ranking permutation.

\begin{table}[!t]
\centering
\caption{Summary of the base models and external search APIs utilized in our proposed framework.}
\label{tab:model_versions}
\resizebox{\columnwidth}{!}{
\begin{tabular}{l l}
\toprule
\textbf{Model / Tool Name} & \textbf{Official Link / Description} \\
\midrule
\multicolumn{2}{c}{\textit{Base Models for \modelname}} \\
\midrule
Qwen3-0.6B & huggingface.co/Qwen/Qwen3-0.6B \\
Qwen3-1.7B & huggingface.co/Qwen/Qwen3-1.7B \\
Qwen3-4B & huggingface.co/Qwen/Qwen3-4B \\
Qwen3-8B & huggingface.co/Qwen/Qwen3-8B \\
Qwen3-14B & huggingface.co/Qwen/Qwen3-14B \\
\midrule
\multicolumn{2}{c}{\textit{External Search APIs \& Tools}} \\
\midrule
Perplexity Pro Search & perplexity.ai \\
\bottomrule
\end{tabular}
}
\end{table}

\section{Training Hyperparameters}
\label{app:implementation}

All experiments are conducted on a single compute node equipped with 8 NVIDIA GPUs. We implement the distributed training pipeline utilizing the DeepSpeed framework. Across all stages, we perform full-parameter fine-tuning rather than Low-Rank Adaptation (LoRA) to maximize the reasoning capacity of the models. The comprehensive hyperparameter settings for both the initial Language-Preserved Warmup (Stage 1) and the subsequent Cost-Aware GRPO (Stage 2) are summarized in Table~\ref{tab:hyperparameters}.
The code will be organized and open-sourced as soon as possible.

\begin{table}[t!]
\centering
\caption{Training hyperparameters for the two-stage pipeline.}
\label{tab:hyperparameters}
\resizebox{\columnwidth}{!}{
\begin{tabular}{l l}
\toprule
\textbf{Setting} & \textbf{Value} \\
\midrule
\multicolumn{2}{c}{\textit{General Infrastructure}} \\
\midrule
Hardware & 8 $\times$ NVIDIA GPUs \\
Distributed Framework & DeepSpeed Zero-3\\
Tuning Strategy & Full-Parameter Fine-Tuning \\
\midrule
\multicolumn{2}{c}{\textit{Stage 1: Language-Preserved Warmup}} \\
\midrule
Training Dataset & MS MARCO \\
Training Epochs & 1 \\
Regularization $\beta$ & 0.01 \\
Optimization Objective & BCE + KL Divergence \\
Learning Rate & $1 \times 10^{-6}$ \\
Per-Device Batch Size & 2 \\
Global Batch Size & 16 \\
Gradient Checkpointing & Disabled \\
\midrule
\multicolumn{2}{c}{\textit{Stage 2: Cost-Aware GRPO}} \\
\midrule
Training Dataset & Sampled MS MARCO \\
Training Epochs & 1 \\
Penalty Intensity $\lambda$ & 0.02 \\
Threshold $\gamma$ & 0.2 \\
Optimization Objective & Designed Reward (MAE) \\
Learning Rate & $5 \times 10^{-7}$ \\
Per-Device Batch Size & 2 \\
Global Batch Size & 16 \\
Gradient Checkpointing & Enabled \\
\bottomrule
\end{tabular}
}
\end{table}
\section{Model Versions and Tool APIs}
\label{app:model_versions}

Throughout this paper, the extensive reranking experiments and scaling analyses are conducted utilizing the specific base model checkpoints listed in Table~\ref{tab:model_versions}. We strictly adhere to the default tokenizer configurations defined in their respective official repositories. The base weights of the Qwen3 family serve as the foundation for training our proposed \modelname~framework. Furthermore, the external search and retrieval tools invoked during the inference phase are also detailed in Table~\ref{tab:model_versions}. Since tool calls are expensive and the test sets contain too many documents, it is computationally infeasible to enforce tool invocation on every query-document pair, making this baseline impractical.

\section{Supplementary Ablation}
\label{app:hyperparameter_ablation}
\begin{figure}[t!]
    \centering
    \includegraphics[width=\columnwidth]{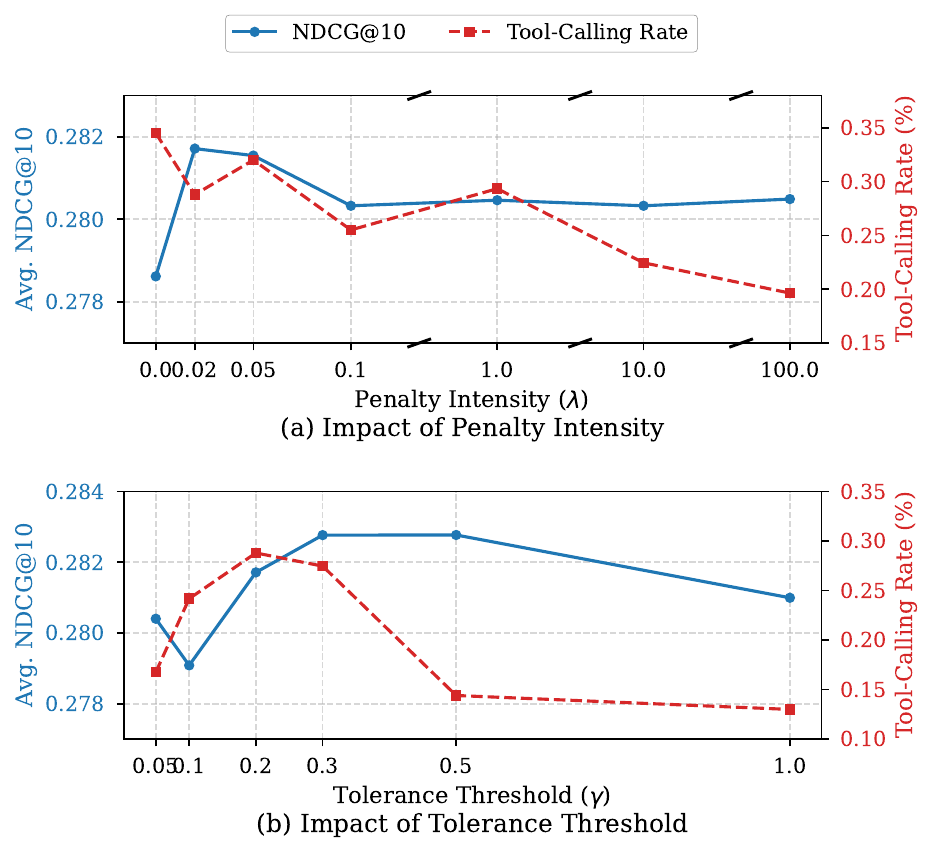}\vspace{-4mm}
    \caption{Ablation study on the penalty intensity $\lambda$ and the tolerance threshold $\gamma$. The left y-axis shows the average NDCG@10 across the BRIGHT benchmark, while the right y-axis represents the overall tool-calling rate.}
    \label{fig:ablation_params}
\end{figure}

To deeply understand the impact of the cost-aware reward on the adaptive routing behavior, we systematically ablate its two core hyperparameters: the latency penalty intensity $\lambda$ and the tolerance threshold $\gamma$. We evaluate the \modelname-4B variant across all BRIGHT subsets. Figure~\ref{fig:ablation_params} presents the detailed results.
As shown in Figure~\ref{fig:ablation_params}(a), increasing the penalty $\lambda$ strictly decreases the tool-calling rate, demonstrating that the model actively learns to suppress external search when the cost is high. However, excessively high penalties ($\lambda \ge 0.1$) force the model to guess blindly, slightly degrading NDCG. A moderate penalty ($\lambda=0.02$) achieves the optimal tradeoff. 
Figure~\ref{fig:ablation_params}(b) illustrates the impact of the tolerance threshold $\gamma$. A threshold that is too strict ($\gamma \le 0.1$) prevents the model from trusting its internal knowledge, whereas a threshold that is too relaxed ($\gamma \ge 0.5$) causes overconfidence and a severe drop in tool usage. Setting $\gamma$ between $0.2$ and $0.3$ optimally balances internal confidence and the necessity for external validation, maximizing the overall accuracy.

To isolate the marginal contribution of tool calling, we compare \modelname-4B and \modelname-8B under two inference modes: \textit{with-tools} and \textit{forced zero-hop} (set \textcolor{teal}{\texttt{<reasoning>}}
\small \textbackslash{}n\textbackslash{}n
\textcolor{teal}{\texttt{</reasoning>}}\textcolor{red}{\texttt{<answer>}} as the suffix). Table~\ref{tab:within_model_ablation} reports the average NDCG@10 across BEIR and BRIGHT.
Tool calling yields meaningful improvements in both benchmarks, demonstrating that tool calls acquire external knowledge and thus accurately improve target accuracy.

\begin{table}[!t]
\centering
\caption{NDCG@10 of \modelname~with adaptive tool calling vs. forced zero-hop. $\Delta$ denotes the gain from tool availability.}\vspace{-2mm}
\label{tab:within_model_ablation}
\resizebox{\columnwidth}{!}{
\begin{tabular}{l c c c c c c c}
\toprule
\multirow{2}{*}{\textbf{Benchmark}} & \multicolumn{3}{c}{\textbf{\modelname-4B}} & \multicolumn{3}{c}{\textbf{\modelname-8B}} \\
\cmidrule(lr){2-4} \cmidrule(lr){5-7}
& w/ tools & zero-hop & $\Delta$ & w/ tools & zero-hop & $\Delta$ \\
\midrule
BEIR Avg. & 46.0 & 42.9 & +3.1 & 46.3 & 43.5 & +2.8 \\
BRIGHT Avg. & 28.2 & 26.1 & +2.1 & 29.7 & 26.8 & +2.9 \\
\bottomrule
\end{tabular}
}\vspace{-4mm}
\end{table}

\section{Case Studies on \modelname}
\label{app:case_studies}

This section provides concrete examples of the inference trajectories generated by \modelname-4B.
In the benchmarks we tested, there were no cases where the number of hops was greater than or equal to 2. These cases highlight the model's ability to selectively invoke external tools based on its parametric confidence, a behavior learned through our cost-aware reinforcement learning phase.

\subsection{Case 1: Resolving Epistemic Uncertainty via Tool Calling (1-Hop)}
In Figure~\ref{fig:case_study_1}, the user's query contains a flawed premise (that Venus is closer to the Sun than Mercury). While the document provides clues (Venus is the second planet and has a CO2 atmosphere), it does not explicitly state the absolute temperatures required to definitively judge the premise. Recognizing this gap in its parametric confidence, \modelname~suspends its scoring process, actively queries the external search engine to retrieve the exact temperatures, and subsequently integrates this feedback to conclude that the document is relevant for correcting the misconception.

\begin{figure}[h]
\centering
\begin{tcolorbox}[colback=gray!5, colframe=gray!40, boxrule=0.5pt, arc=4pt, left=6pt, right=6pt, top=6pt, bottom=6pt]

\textbf{[User Query]} Is the average surface temperature of Venus higher than Mercury because of its closer proximity to the Sun? \\
\textbf{[Candidate Document]} \textit{...Mercury is the closest planet to the Sun at a distance of 58 million km. Venus is the second planet from the Sun. Venus has a thick atmosphere composed of 96\% carbon dioxide...}
\\
\textbf{[Suffix]}
\textcolor{teal}{\texttt{<reasoning>}}
\small \textbackslash{}n\textbackslash{}n
\textcolor{teal}{\texttt{</reasoning>}}

\vspace{1mm}
\hrule
\vspace{2mm}

\textbf{\textcolor{blue}{[\modelname~Generation Starts]}}

\textcolor{orange}{\texttt{<tool\_call>}} \\
\small \texttt{\{"name": "perplexity", "arguments": \{"query\_list": ["average surface temperature Venus vs Mercury"]\}\}} \\
\textcolor{orange}{\texttt{</tool\_call>}}

\vspace{1mm}
\hrule
\vspace{2mm}

\textbf{\textcolor{gray}{[Environment Execution]}} \\
\textcolor{gray}{\texttt{<tool\_response>}} \\
\small Venus has an average surface temperature of 464 °C, making it the hottest planet. Mercury's average temperature is 167 °C, though daytimes reach 430 °C. \\
\textcolor{gray}{\texttt{</tool\_response>}}

\vspace{1mm}
\hrule
\vspace{2mm}

\textbf{\textcolor{blue}{[\modelname~Generation Resumes]}}

\textcolor{teal}{\texttt{<reasoning>}} \\
\small The tool confirms Venus is hotter (464 °C vs 167 °C). The document explains why: it mentions Venus's thick CO2 atmosphere and corrects the distance premise. Therefore, the document provides the exact facts needed to directly resolve the user's misconception. It is highly relevant. \\
\textcolor{teal}{\texttt{</reasoning>}}

\textcolor{red}{\texttt{<answer>}} \\
\small \texttt{yes} \\
\textcolor{red}{\texttt{</answer>}}

\end{tcolorbox}
\caption{A 1-Hop inference trajectory. The model utilizes the \texttt{<reasoning>} tag for internal deliberation, dynamically invokes a \texttt{<tool\_call>} to resolve epistemic uncertainty regarding planetary temperatures, and seamlessly resumes reasoning to emit the final terminal \texttt{<answer>}.}
\label{fig:case_study_1}
\end{figure}

\subsection{Case 2: Direct Scoring via Internal Knowledge (0-Hop)}
Figure~\ref{fig:case_study_2} illustrates a contrasting scenario where the model evaluates a query regarding fundamental physics. The candidate document explicitly lists the four fundamental forces. Because this information is robustly grounded within the model's pre-trained parametric memory, it exhibits high confidence during its internal \texttt{<reasoning>} step. Consequently, the model correctly bypasses the \texttt{<tool\_call>} action entirely, avoiding unnecessary latency, and directly emits the terminal \texttt{<answer>}.

\begin{figure}[h]
\centering
\begin{tcolorbox}[colback=gray!5, colframe=gray!40, boxrule=0.5pt, arc=4pt, left=6pt, right=6pt, top=6pt, bottom=6pt]

\textbf{[User Query]} What are the four fundamental forces of nature that govern the universe? \\
\textbf{[Candidate Document]} \textit{...The universe is governed by four fundamental forces: gravity, electromagnetism, the strong nuclear force, and the weak nuclear force. Gravity is the weakest but has an infinite range, while the strong force binds atomic nuclei...}  \\
\textbf{[Suffix]}
\textcolor{teal}{\texttt{<reasoning>}}
\small \textbackslash{}n\textbackslash{}n
\textcolor{teal}{\texttt{</reasoning>}}

\vspace{1mm}
\hrule
\vspace{2mm}

\textbf{\textcolor{blue}{[\modelname~Generation Starts]}}

\textcolor{red}{\texttt{<answer>}} \\
\small \texttt{yes} \\
\textcolor{red}{\texttt{</answer>}}

\end{tcolorbox}
\caption{A 0-Hop inference trajectory. The model confidently evaluates a fundamental physics query within its parametric boundaries, bypassing the \texttt{<tool\_call>} action to emit an immediate \texttt{<answer>} and minimizing latency.}
\label{fig:case_study_2}
\end{figure}


\end{document}